\documentclass[runningheads]{llncs}
\usepackage{graphicx}
\usepackage{amssymb}
\usepackage{hyperref}
\usepackage[table]{xcolor}
\usepackage{booktabs}
\usepackage{caption}
\usepackage{subcaption}
\usepackage{sidecap}

\usepackage[color=red]{todonotes}
\usepackage{marginnote}

\newcolumntype{?}{!{\vrule width 1pt}}

\begin{document}
\title{Understanding Silent Failures \\ in Medical Image Classification}
\titlerunning{Silent Failures in Medical Image Classification}
%
\author{Till J. Bungert\inst{1,2} \and
Levin Kobelke\inst{1,2} \and
Paul F. Jaeger\inst{1,2}}

\authorrunning{T. Bungert et al.}
\institute{Interactive Machine Learning Group, German Cancer Research Center (DKFZ), Heidelberg, Germany \and
Helmholtz Imaging, DKFZ, Heidelberg, Germany
\email{till.bungert@dkfz-heidelberg.de}}

\maketitle              
\begin{abstract}
    To ensure the reliable use of classification systems in medical applications, it is crucial to prevent silent
    failures. This can be achieved by either designing classifiers that are robust enough to avoid failures in
    the first place, or by detecting remaining failures using confidence scoring functions (CSFs). A predominant
    source of failures in image classification is distribution shifts between training data and deployment data.
    To understand the current state of silent failure prevention in medical imaging, we conduct the first
    comprehensive analysis comparing various CSFs in four biomedical tasks and a diverse range of distribution
    shifts. Based on the result that none of the benchmarked CSFs can reliably prevent silent failures, we
    conclude that a deeper understanding of the root causes of failures in the data is required. To facilitate
    this, we introduce SF-Visuals, an interactive analysis tool that uses latent space clustering to visualize
    shifts and failures. On the basis of various examples, we demonstrate how this tool can help researchers gain
    insight into the requirements for safe application of classification systems in the medical domain. The
    open-source benchmark and tool are at: \url{https://github.com/IML-DKFZ/sf-visuals}.

\end{abstract}

\section{Introduction}%
    \label{sec:introduction}
    \begin{figure}[h!]
        \begin{center}
            \includegraphics[width=1\textwidth]{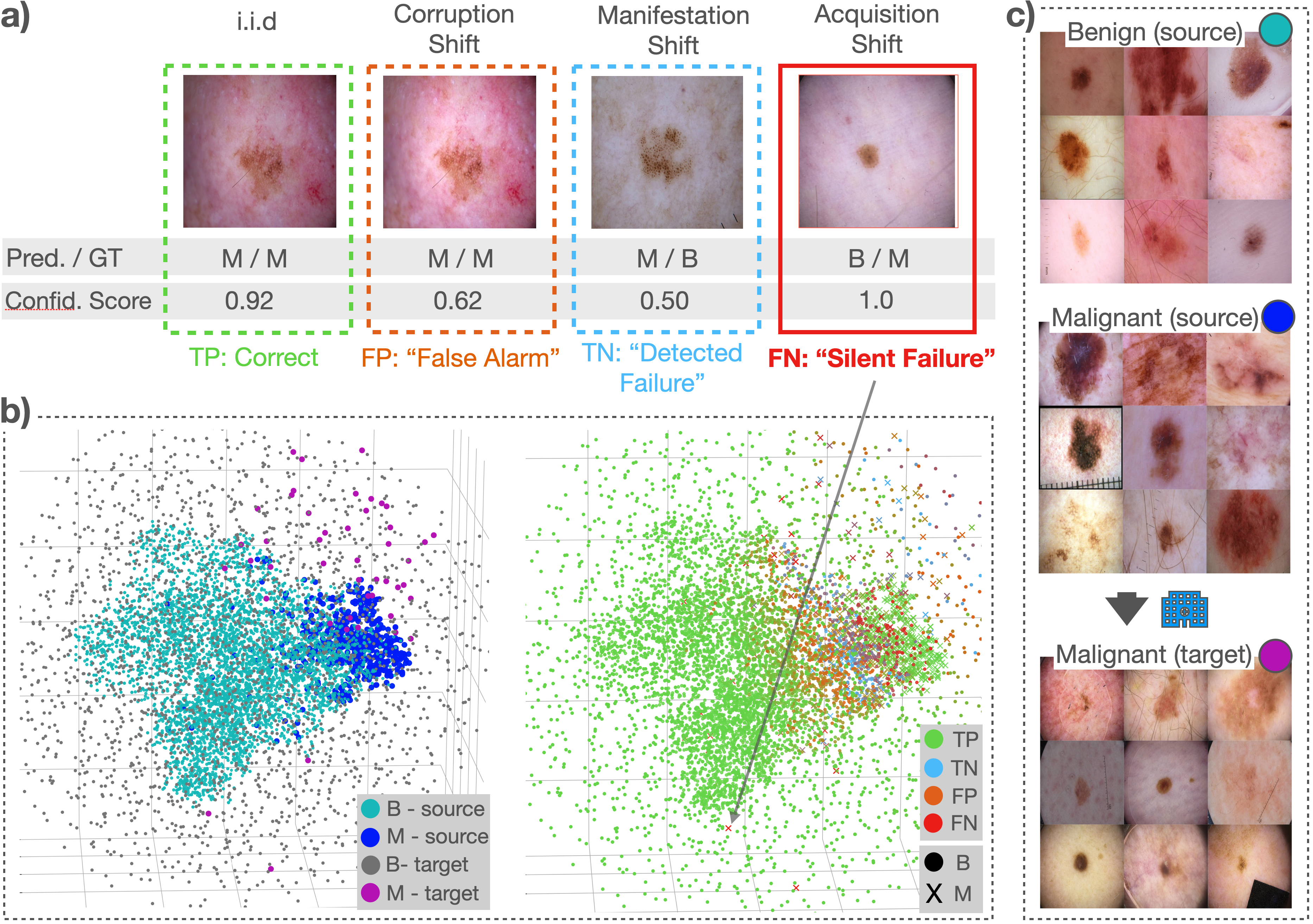}
        \end{center}
        \caption{\textbf{a)} Exemplary predictions of the classifier and the accompanying confidence scoring function (CSF, here: ConfidNet) on the dermoscopy dataset across several distribution shifts. \textbf{Note that True/False Positives/Negatives (T/F P/N) do not refer to the classifier decision, but to the failure detection outcome, i.e. the assessment of the CSF.} In this context, FN, i.e. cases with incorrect predictions ("failure") and a high confidence score  ("failure not detected") are referred to as \textit{silent failures}. \textbf{b)} SF-Visuals allows to identify and analyze silent failures in a dataset based on an Interactive Scatter Plot in the classifier's latent space (each dot represents one image, which is displayed when selecting the dot). \textbf{c)} SF-Visuals further features Concept Cluster Plots to gain an intuition of how the model perceives distinct classes or distribution shifts. More details on the displayed example are in Section~\ref{sec:investigation_of_failure sources}. Abbreviations: B: Benign, M: Malignant, Pred.: Prediction, GT: Ground truth, Confid.: Confidence, Source: Source domain, Target: Target domain.}
        \label{fig:overview}
    \end{figure}
    Although machine learning-based classification systems have achieved significant breakthroughs in various
    research and practical areas, their clinical application is still lacking. A primary reason is the lack of
    reliability, i.e. failure cases produced by the system, which predominantly occur when deployment data
    differs from the data it was trained on, a phenomenon known as \textit{distribution shifts}. In medical
    applications, these shifts can be caused by image corruption (“corruption shift”), unseen variants of
    pathologies (“manifestation shift”), or deployment in new clinical sites with different scanners and
    protocols (“acquisition shift”) \cite{castroCausalityMattersMedical2020a}. The \textit{robustness} of a
    classifier, i.e. its ability to generalize across these shifts, is extensively studied in the computer vision
    community with a variety of recent benchmarks covering nuanced realistic distribution shifts
    \cite{idrissiImageNetXUnderstandingModel2022,kohWILDSBenchmarkIntheWild2021,santurkarBREEDSBenchmarksSubpopulation2022,jaegerCallReflectEvaluation2022a},
    and is also studied in isolated cases in the biomedical community
    \cite{zhangBenchmarkingRobustnessDeep2022,bandBenchmarkingBayesianDeep2022,bernhardtFailureDetectionMedical2022}.
    Despite these efforts, perfect classifiers are not to be expected, thus a second mitigation strategy is to
    detect and defer the remaining failures, thus \textit{preventing failures to be silent}. This is done by
    means of confidence scoring functions (CSF) of different types as studied in the fields of misclassification
    detection (MisD)
    \cite{corbiereAddressingFailurePrediction2019a,hendrycksBaselineDetectingMisclassified2018,malininPredictiveUncertaintyEstimation2018},
    Out-of-Distribution detection (OoD-D)
    \cite{fortExploringLimitsOutofDistribution2021,winkensContrastiveTrainingImproved2020,leeSimpleUnifiedFramework2018,hendrycksBaselineDetectingMisclassified2018,devriesLearningConfidenceOutofDistribution2018a,liangEnhancingReliabilityOutofdistribution2020},
    selective classification (SC)
    \cite{liuDeepGamblersLearning2019,geifmanSelectiveClassificationDeep2017,geifmanSelectiveNetDeepNeural2019},
    and predictive uncertainty quantification (PUQ) \cite{ovadiaCanYouTrust2019,kendallWhatUncertaintiesWe2017}.

    \textit{We argue, that silent failures, which occur when test cases break both the classifier and the CSF, are a significant bottleneck in the clinical translation of ML systems and require further attention in the medical community.}

    Note that the task of silent failure prevention is orthogonal to calibration, as, for example, a perfectly calibrated classifier can still yield substantial amounts of silent failures and vice versa~\cite{jaegerCallReflectEvaluation2022a}.

    Bernhardt et al.\cite{bernhardtFailureDetectionMedical2022} studied failure detection on several biomedical
    datasets, but only assessed the performance of CSFs in isolation without considering the classifier's ability
    to prevent failures. Moreover, their study did not include distribution shifts thus lacking a wide range of
    realistic failure sources. Jaeger et al.\cite{jaegerCallReflectEvaluation2022a}, on the other hand, recently
    discussed various shortcomings in current research on silent failures including the common lack of
    distribution shifts and the lack of assessing the classifier and CSF as a joint system. However, their study
    did not cover tasks from the biomedical domain.

    In this work, our contribution is twofold: \textbf{1)} Building on the work of Jaeger et
    al.\cite{jaegerCallReflectEvaluation2022a}, we present the first comprehensive study of silent failure
    prevention in the biomedical field. We compare various CSFs under a wide range of distribution shifts on four
    biomedical datasets. Our study provides valuable insights and the underlying framework is made openly
    available to catalyze future research in the community. \textbf{2)} Since the benchmark reveals that none of
    the predominant CSFs can reliably prevent silent failures in biomedical tasks, we argue that a deeper
    understanding of the root causes in the data itself is required. To this end, we present SF-Visuals, a
    visualization tool that facilitates identifying silent failures in a dataset and investigating their causes
    (see Figure~\ref{fig:overview}). Our approach contributes to recent research on visual analysis of failures
    \cite{idrissiImageNetXUnderstandingModel2022}, which has not focused on silent failures and distribution
    shifts before.

\section{Methods} 
    \label{sec:methods}
    \textbf{Benchmark for Silent Failure Prevention under Distribution Shifts.}
    We follow the spirit of recent robustness benchmarks, where existing datasets have been enhanced by various distribution shifts to evaluate methods under a wide range of failure sources and thus simulate real-world application~\cite{santurkarBREEDSBenchmarksSubpopulation2022,kohWILDSBenchmarkIntheWild2021}. To our knowledge, no such comprehensive benchmark currently exists in the biomedical domain. Specifically, we introduce corruptions of various intensity levels to the images in four datasets in the form of brightness, motion blur, elastic transformations and Gaussian noise. We further simulate acquisition shifts and manifestation shifts by splitting the data into "source domain" (development data) and "target domain" (deployment data) according to sub-class information from the meta-data such as lesion subtypes or clinical sites. \textbf{Dermoscopy dataset:} We combine data from ISIC 2020~\cite{rotembergPatientcentricDatasetImages2021}, derma 7 point~\cite{kawaharaSevenPointChecklistSkin2019}, PH2~\cite{mendoncaPH2DermoscopicImage2013} and HAM10000~\cite{tschandlHAM10000DatasetLarge2018} and map all lesion sub-types to the super-classes “benign” or “malignant”. We emulate two acquisition shifts by defining either images from the Memorial Sloan Kettering Cancer Center (MSKCC) or Hospital Clinic Barcelona (HCB) as the target domain and the remaining images as the source domain. Further, a manifestation shift is designed by defining the lesion subtypes "keratosis-like" (benign) and "actinic keratosis" (malignant) as the target domain. \textbf{Chest X-ray dataset:} We pool the data from CheXpert~\cite{irvinCheXpertLargeChest2019}, NIH14~\cite{wangChestXRay8HospitalScaleChest2017} and MIMIC~\cite{johnsonMIMICIIIFreelyAccessible2016}, while only retaining the classes common to all three. Next, we emulate two acquisition shifts by defining either the NIH14 or the CheXpert data as the target domain. \textbf{FC-Microscopy dataset:} The RxRx1 dataset~\cite{sypetkowskiRxRx1DatasetEvaluating2023} represents the fluorescence cell microscopy domain. Since the images were acquired in 51 deviating acquisition steps, we define 10 of these batches as target-domain to emulate an acquisition shift. \textbf{Lung Nodule CT dataset:} We create a simple 2D binary nodule classification task based on the 3D LIDC-IDRI data~\cite{armatoLungImageDatabase2011} by selecting the slice with the largest annotation per nodule ($\pm$ two slices resulting in 5 slices per nodule). Average malignancy ratings (four raters per nodule, scores between 1 and 5)  $>2$ are considered malignant and all others as benign. We emulate two manifestation shifts by defining nodules with high spiculation (rating $>$2), and low texture (rating $<$3) as target domains.\\
    The datasets consist only of publicly available data, our benchmark provides scripts to automatically
    generate the combined datasets and distribution shifts.\\\\
    \textbf{The SF-Visuals Tool: Visualizing Silent Failures.}
    The proposed tool is based on three simple operations, that enable effective and intuitive analysis of silent failures in datasets across various CSFs: \textit{1) Interactive Scatter Plots:} See example in Figure~\ref{fig:overview}b. We first reduce the dimensionality of the classifier’s latent space to 50 using principal component analysis and use t-SNE to obtain the final 3-dimensional embedding. Interactive functionality includes coloring dots via pre-defined schemes such as classes, distribution shifts, classifier confusion matrix, or CSF confusion matrix. The associated images are displayed upon selection of a dot to establish a direct visual link between input space and embedding. \textit{2) Concept Cluster Plots:} See examples in Figure~\ref{fig:overview}c. To abstract away from individual points in the scatter plot, concepts of interest, such as classes or distribution shifts can be defined and visualized to identify conceptual commonalities and differences in the data as perceived by the model. Therefore, k-means clustering is applied to the 3-dimensional embedding. Nine clusters are identified per concept and the resulting plots show the closest-to-center image per cluster as a visual representation of the concept. \textit{3) Silent Failure Visualization:} See examples in Figure~\ref{fig:bottom_up}. We sort all failures by the classifier confidence and by default show the images associated with the top-two most confident failures. For corruption shifts, we further allow investigating the predictions on a fixed input image over varying intensity levels.\\\\
    Based on these visualizations, the functionality of SF-Visuals is three-fold: 1) Visual analysis of the
    dataset including distribution shifts. 2) Visual analysis of the general behavior of various CSFs on a given task
    3) Visual analysis of individual silent failures in the dataset for various CSFs.
\section{Experimental Setup} 
    \label{sec:setup}
    \textbf{Evaluating Silent Failure Prevention:} 
    We follow Jaeger et al.~\cite{jaegerCallReflectEvaluation2022a} in evaluating silent failure prevention as a joint task of the classifier and the CSF. The area under the risk-coverage curve AURC reflects this task, since it considers both the classifier's accuracy as well as the CSF's ability to detect failures by assigning low confidence scores. Thus, it can be interpreted as a \textit{silent failure rate} or the error rate averaged over steps of filtering cases one by one according to their rank of confidence score (low to high). Exemplary risk-coverage curves are shown in Appendix Figure~3. \textbf{Compared Confidence Scoring Functions:} We compare the following CSFs: The maximum softmax response (MSR) and the predictive entropy computed from the classifier's softmax output, three predictive uncertainty measures based on Monte-Carlo Dropout (MCD)~\cite{galDropoutBayesianApproximation2016}, namely mean softmax (MCD-MSR), predictive entropy (MCD-PE) and expected entropy (MCD-EE), ConfidNet~\cite{corbiereAddressingFailurePrediction2019a}, which is trained as an extension to the classifier, DeepGamblers (DG) that learns a confidence like reservation score (DG-Res)~\cite{liuDeepGamblersLearning2019} and the work of DeVries et al.~\cite{devriesLearningConfidenceOutofDistribution2018a}. \textbf{Training Settings:} On each dataset, we employ the classifier behind the respective leading results in literature: For chest X-ray data we use DenseNet121~\cite{huangDenselyConnectedConvolutional2017}, for dermoscopy data we use EfficientNet-B4~\cite{tanEfficientNetRethinkingModel2019} and for fluorescence cell microscopy and lung nodule CT data we us DenseNet161~\cite{huangDenselyConnectedConvolutional2017}. We select the initial learning rate between $10^{-3}$ and $10^{-5}$ and weight decay between 0 and $10^{-5}$ via grid search and optimize for validation accuracy. All models were trained with dropout. All hyperparameters can be found in Appendix Table~3.
\section{Results} 
    \label{sub:results}

    \subsection{Silent Failure Prevention Benchmark} 
        \label{sub:silent_failure_detection_benchmark}
        \begin{table}[t]
            \begin{center}
                \resizebox{\textwidth}{!}{                 \begin{tabular}{l?rrr?rrrr?rrr?rrr}
Dataset & \multicolumn{3}{c}{Chest X-ray} & \multicolumn{4}{c}{Dermoscopy} & \multicolumn{3}{c}{FC-Microscopy} & \multicolumn{3}{c}{Lung Nodule CT} \\
Study & iid & cor & acq & iid & cor & acq & man & iid & cor & acq & iid & cor & man \\
\midrule
MSR & {\cellcolor[HTML]{FEDCBB}} \color[HTML]{000000} 15.3 & {\cellcolor[HTML]{FEECD9}} \color[HTML]{000000} 18.6 & {\cellcolor[HTML]{FEEAD6}} \color[HTML]{000000} 23.1 & {\cellcolor[HTML]{FFF5EB}} \color[HTML]{000000} 0.544 & {\cellcolor[HTML]{FFF1E3}} \color[HTML]{000000} 0.913 & {\cellcolor[HTML]{FFF3E6}} \color[HTML]{000000} 0.799 & {\cellcolor[HTML]{E75B0B}} \color[HTML]{F1F1F1} 49.3 & {\cellcolor[HTML]{FDA660}} \color[HTML]{000000} 13.3 & {\cellcolor[HTML]{FEE4CA}} \color[HTML]{000000} 55.6 & {\cellcolor[HTML]{FFF3E6}} \color[HTML]{000000} 32.4 & {\cellcolor[HTML]{7F2704}} \color[HTML]{F1F1F1} 6.69 & {\cellcolor[HTML]{FD984B}} \color[HTML]{000000} 8.18 & {\cellcolor[HTML]{FEEDDC}} \color[HTML]{000000} 12.1 \\
PE & {\cellcolor[HTML]{FDD5AD}} \color[HTML]{000000} 15.5 & {\cellcolor[HTML]{FEE7D1}} \color[HTML]{000000} 18.9 & {\cellcolor[HTML]{FEE4CA}} \color[HTML]{000000} 23.6 & {\cellcolor[HTML]{FFF5EB}} \color[HTML]{000000} 0.544 & {\cellcolor[HTML]{FFF1E3}} \color[HTML]{000000} 0.913 & {\cellcolor[HTML]{FFF3E6}} \color[HTML]{000000} 0.799 & {\cellcolor[HTML]{E75B0B}} \color[HTML]{F1F1F1} 49.3 & {\cellcolor[HTML]{FD984B}} \color[HTML]{000000} 14.1 & {\cellcolor[HTML]{FDD7AF}} \color[HTML]{000000} 56.3 & {\cellcolor[HTML]{FFF2E5}} \color[HTML]{000000} 32.7 & {\cellcolor[HTML]{7F2704}} \color[HTML]{F1F1F1} 6.69 & {\cellcolor[HTML]{FD984B}} \color[HTML]{000000} 8.18 & {\cellcolor[HTML]{FEEDDC}} \color[HTML]{000000} 12.1 \\
MCD-MSR & {\cellcolor[HTML]{FEE9D4}} \color[HTML]{000000} 14.9 & {\cellcolor[HTML]{FFF5EB}} \color[HTML]{000000} 17.9 & {\cellcolor[HTML]{FFF5EB}} \color[HTML]{000000} 22.1 & {\cellcolor[HTML]{FFF5EB}} \color[HTML]{000000} 0.544 & {\cellcolor[HTML]{FFF1E3}} \color[HTML]{000000} 0.913 & {\cellcolor[HTML]{FFF3E6}} \color[HTML]{000000} 0.799 & {\cellcolor[HTML]{E75B0B}} \color[HTML]{F1F1F1} 49.3 & {\cellcolor[HTML]{FDB373}} \color[HTML]{000000} 12.6 & {\cellcolor[HTML]{FDD3A9}} \color[HTML]{000000} 56.5 & {\cellcolor[HTML]{FFF5EB}} \color[HTML]{000000} 31.8 & {\cellcolor[HTML]{E75C0C}} \color[HTML]{F1F1F1} 5.80 & {\cellcolor[HTML]{FFF5EB}} \color[HTML]{000000} 7.13 & {\cellcolor[HTML]{FFF5EB}} \color[HTML]{000000} 11.5 \\
MCD-PE & {\cellcolor[HTML]{FEE4CA}} \color[HTML]{000000} 15.1 & {\cellcolor[HTML]{FFF1E4}} \color[HTML]{000000} 18.2 & {\cellcolor[HTML]{FFEEDE}} \color[HTML]{000000} 22.7 & {\cellcolor[HTML]{FFF5EB}} \color[HTML]{000000} 0.544 & {\cellcolor[HTML]{FFF1E3}} \color[HTML]{000000} 0.913 & {\cellcolor[HTML]{FFF3E6}} \color[HTML]{000000} 0.799 & {\cellcolor[HTML]{E75B0B}} \color[HTML]{F1F1F1} 49.3 & {\cellcolor[HTML]{FDA863}} \color[HTML]{000000} 13.2 & {\cellcolor[HTML]{FDC189}} \color[HTML]{000000} 57.2 & {\cellcolor[HTML]{FFF4E9}} \color[HTML]{000000} 32.1 & {\cellcolor[HTML]{E75C0C}} \color[HTML]{F1F1F1} 5.80 & {\cellcolor[HTML]{FFF5EB}} \color[HTML]{000000} 7.13 & {\cellcolor[HTML]{FFF5EB}} \color[HTML]{000000} 11.5 \\
MCD-EE & {\cellcolor[HTML]{FEE4CA}} \color[HTML]{000000} 15.1 & {\cellcolor[HTML]{FFF1E4}} \color[HTML]{000000} 18.2 & {\cellcolor[HTML]{FFEEDE}} \color[HTML]{000000} 22.7 & {\cellcolor[HTML]{FFF5EB}} \color[HTML]{000000} 0.544 & {\cellcolor[HTML]{FFF1E3}} \color[HTML]{000000} 0.913 & {\cellcolor[HTML]{FFF3E6}} \color[HTML]{000000} 0.799 & {\cellcolor[HTML]{E75B0B}} \color[HTML]{F1F1F1} 49.3 & {\cellcolor[HTML]{FDA660}} \color[HTML]{000000} 13.3 & {\cellcolor[HTML]{FDC189}} \color[HTML]{000000} 57.2 & {\cellcolor[HTML]{FFF4E9}} \color[HTML]{000000} 32.1 & {\cellcolor[HTML]{F16813}} \color[HTML]{F1F1F1} 5.68 & {\cellcolor[HTML]{FFF4E8}} \color[HTML]{000000} 7.16 & {\cellcolor[HTML]{FFF0E1}} \color[HTML]{000000} 11.9 \\
ConfidNet & {\cellcolor[HTML]{FEE4CA}} \color[HTML]{000000} 15.1 & {\cellcolor[HTML]{FEEDDC}} \color[HTML]{000000} 18.5 & {\cellcolor[HTML]{FEEDDC}} \color[HTML]{000000} 22.8 & {\cellcolor[HTML]{FEE4CA}} \color[HTML]{000000} 0.581 & {\cellcolor[HTML]{FEE2C6}} \color[HTML]{000000} 0.979 & {\cellcolor[HTML]{FFF1E3}} \color[HTML]{000000} 0.806 & {\cellcolor[HTML]{7F2704}} \color[HTML]{F1F1F1} 51.1 & {\cellcolor[HTML]{7F2704}} \color[HTML]{F1F1F1} 21.9 & {\cellcolor[HTML]{7F2704}} \color[HTML]{F1F1F1} 63.7 & {\cellcolor[HTML]{7F2704}} \color[HTML]{F1F1F1} 61.9 & {\cellcolor[HTML]{EA5F0E}} \color[HTML]{F1F1F1} 5.77 & {\cellcolor[HTML]{FEE0C1}} \color[HTML]{000000} 7.50 & {\cellcolor[HTML]{FD9243}} \color[HTML]{000000} 15.7 \\
DG-MCD-MSR & {\cellcolor[HTML]{FFF5EB}} \color[HTML]{000000} 14.4 & {\cellcolor[HTML]{FEE6CE}} \color[HTML]{000000} 19.0 & {\cellcolor[HTML]{FDD6AE}} \color[HTML]{000000} 24.4 & {\cellcolor[HTML]{FDD1A3}} \color[HTML]{000000} 0.611 & {\cellcolor[HTML]{FFF5EB}} \color[HTML]{000000} 0.893 & {\cellcolor[HTML]{FFF5EB}} \color[HTML]{000000} 0.787 & {\cellcolor[HTML]{BB3D02}} \color[HTML]{F1F1F1} 50.1 & {\cellcolor[HTML]{FFF5EB}} \color[HTML]{000000} 7.46 & {\cellcolor[HTML]{FFF5EB}} \color[HTML]{000000} 54.3 & {\cellcolor[HTML]{FFF0E1}} \color[HTML]{000000} 33.2 & {\cellcolor[HTML]{FFF5EB}} \color[HTML]{000000} 3.97 & {\cellcolor[HTML]{B53B02}} \color[HTML]{F1F1F1} 9.04 & {\cellcolor[HTML]{FEE0C1}} \color[HTML]{000000} 12.9 \\
DG-RES & {\cellcolor[HTML]{7F2704}} \color[HTML]{F1F1F1} 19.4 & {\cellcolor[HTML]{7F2704}} \color[HTML]{F1F1F1} 26.5 & {\cellcolor[HTML]{7F2704}} \color[HTML]{F1F1F1} 32.8 & {\cellcolor[HTML]{7F2704}} \color[HTML]{F1F1F1} 0.814 & {\cellcolor[HTML]{7F2704}} \color[HTML]{F1F1F1} 1.46 & {\cellcolor[HTML]{7F2704}} \color[HTML]{F1F1F1} 1.32 & {\cellcolor[HTML]{FDD3A9}} \color[HTML]{000000} 46.8 & {\cellcolor[HTML]{FDD6AE}} \color[HTML]{000000} 10.6 & {\cellcolor[HTML]{FEECDA}} \color[HTML]{000000} 55.0 & {\cellcolor[HTML]{FDD7B1}} \color[HTML]{000000} 38.1 & {\cellcolor[HTML]{FDB373}} \color[HTML]{000000} 4.94 & {\cellcolor[HTML]{C54102}} \color[HTML]{F1F1F1} 8.95 & {\cellcolor[HTML]{FDA762}} \color[HTML]{000000} 15.0 \\
Devries et al. & {\cellcolor[HTML]{FFEEDD}} \color[HTML]{000000} 14.7 & {\cellcolor[HTML]{FFEEDE}} \color[HTML]{000000} 18.4 & {\cellcolor[HTML]{FEE5CC}} \color[HTML]{000000} 23.5 & {\cellcolor[HTML]{8E2D04}} \color[HTML]{F1F1F1} 0.801 & {\cellcolor[HTML]{FDBA7F}} \color[HTML]{000000} 1.08 & {\cellcolor[HTML]{FEDDBC}} \color[HTML]{000000} 0.882 & {\cellcolor[HTML]{FFF5EB}} \color[HTML]{000000} 45.5 & {\cellcolor[HTML]{FDAE6A}} \color[HTML]{000000} 12.9 & {\cellcolor[HTML]{B03903}} \color[HTML]{F1F1F1} 62.3 & {\cellcolor[HTML]{EC620F}} \color[HTML]{F1F1F1} 51.4 & {\cellcolor[HTML]{FDAE6A}} \color[HTML]{000000} 4.99 & {\cellcolor[HTML]{7F2704}} \color[HTML]{F1F1F1} 9.41 & {\cellcolor[HTML]{7F2704}} \color[HTML]{F1F1F1} 20.2 \\
\bottomrule
\end{tabular}
                }
            \end{center}
            \caption{\textbf{Silent failure prevention benchmark results measured in $\mathrm{AURC} [\%]]$
            (score range: [0, 100], lower is better).} The coloring is normalized by column, while lighter
            colors depict better scores. All values denote an average of three runs. "cor" denotes the average over
            all corruption types and intensities levels. Similarly, "acq"/"man" denote averages over all acquisition/manifestation shifts per dataset. "iid" denotes scenarios without distribution shifts. Results with further metrics are reported in Appendix Table~2}
            \label{tab:benchmark}
        \end{table}

        Table~\ref{tab:benchmark} shows the results of our benchmark for silent failure prevention in the biomedical
        domain and provides the first overview of the current state of the reliability of classification systems in
        high-stake biomedical applications.

        \textbf{None of the evaluated methods from the literature beats the Maximum Softmax Response baseline across a realistic range of failure sources.} This result is generally consistent with previous findings in Bernhard et al.~\cite{bernhardtFailureDetectionMedical2022} and Jaeger et al.~\cite{jaegerCallReflectEvaluation2022a}, but is shown for the first time for a diverse range of realistic biomedical failure sources. Previously proposed methods do not outperform MSR baselines even in the settings they have been proposed for, e.g. Devries et al. under distribution shifts, or ConfidNet and DG-RES for i.i.d. testing.

        \textbf{MCD and loss attenuation are able to improve the MSR.}
        MCD-MSR is the overall best performing method indicating that MCD generally improves the confidence scoring ability of softmax outputs on these tasks. Interestingly, the DG loss attenuation applied to MCD-MSR, DG-MCD-MSR, which has not been part of the original DG publication but was first tested in Jaeger et al.~\cite{jaegerCallReflectEvaluation2022a}, shows the best results on i.i.d. testing on 3 out of 4 tasks. However, the method is not reliable across all settings, falling short on manifestation shifts and corruptions on the lung nodule CT dataset.

        \textbf{Effects of particular shifts on the reliability of a CSF might be interdependent.}
        When looking beyond the averages displayed in Table~\ref{tab:benchmark} and analyzing the results of individual clinical centers, corruptions and manifestation shifts, one remarkable pattern can be observed: In various cases, the same CSF showed opposing behavior between two variants of the same shift on the same dataset. For instance, Devries et al. outperforms all other CSFs for one clinical site (MSKCC) as target domain, but falls short on the other one (HCB). On the Chest X-ray dataset, MCD worsens the performance for darkening corruptions across all CSFs and intensity levels, whereas the opposite is observed for brightening corruptions. Further, on the lung nodule CT dataset, DG-MCD-RES performs best on bright/dark corruptions and the spiculation manifestation shift, but worst on noise corruption and falls behind on the texture manifestation shift. These observations indicate trade-offs, where, within one distribution shift, reliability against one domain might induce susceptibility to other domains.

        \textbf{Current systems are not generally reliable enough for clinical application.} Although CSFs can mitigate the rate of silent failures (see Appendix Figure~3), the reliability of the resulting classification systems is not sufficient for high-stake applications in the biomedical domain, with substantial rates of silent failure in three out of four tasks. Therefore, a deeper understanding of the root causes of these failures is needed.

    \subsection{Investigation of Silent Failure Sources} 
        \label{sec:investigation_of_failure sources}
        \begin{figure}[h!]
            \begin{center}
                \includegraphics[width=0.92\textwidth]{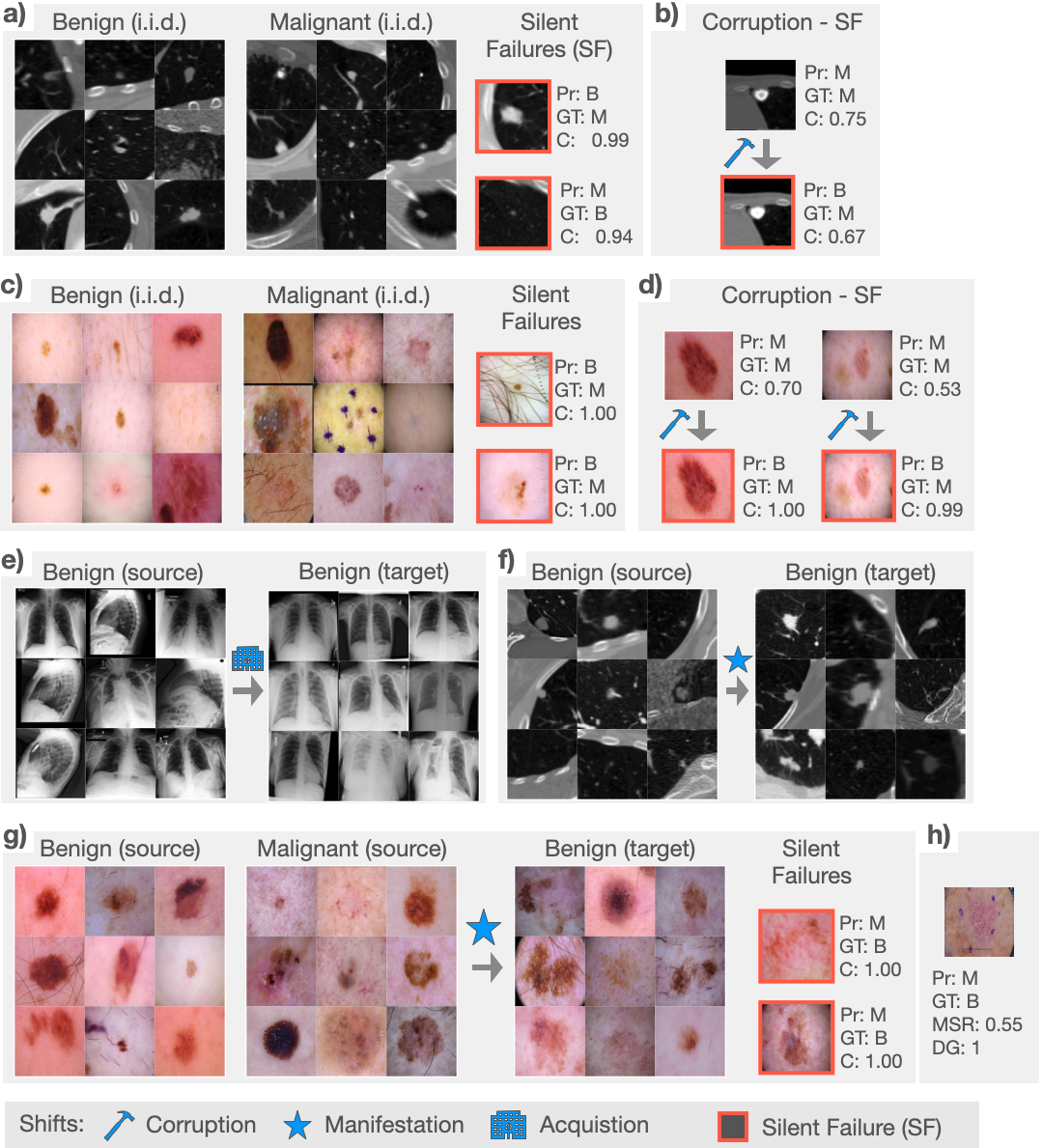}
            \end{center}
            \caption{Various Examples of how the SF-Visuals tool fosters a deeper understanding of root causes of silent failures. Abbreviations: i.i.d: Independent and identically distributed, Pr.: Prediction. GT: Ground Truth, C: Confidence Score, Source: Source domain, Target: Target domain.}
            \label{fig:bottom_up}
        \end{figure}
        \textbf{SF-Visuals enables comprehensive analysis of silent failures} Figure~\ref{fig:overview} vividly demonstrates the added benefit of the proposed tool. First, an Interactive Scatter Plot (Figure~\ref{fig:overview}b, left) provides an overview of the MSKCC acquisition shift on the dermoscopy dataset and reveals a severe change of the data distribution. For instance, some malignant lesions of the target domain (purple dots) are located deep within the "benign" cluster. Figure~\ref{fig:overview}c provides a Concept Cluster Plot that visually confirms how some of these lesions (purple dot) share characteristics of the benign cluster of the source domain (turquoise dot), such as being smaller, brighter, and rounder compared to malignant source-lesions (blue dot). The right-hand plot of Figure~\ref{fig:overview}b reveals that these cases have in fact caused silent failures (red crosses) and visual inspection (see arrow and Figure~\ref{fig:overview}a) confirms the hypothesis that these failures have been caused by the fact that the acquisition shift introduced malignant target-lesions that exhibit benign characteristics.
        Figure~\ref{fig:overview}b (right) further provides insights about the general behavior of the CSF: Silent failures occur for both classes and are either located at the cluster border (i.e. decision boundary), deeper inside the opposing cluster center (severe class confusions), or represent outliers. Most silent failures occur at the boundary, where the CSF should reflect class ambiguities by low scores, hinting at general misbehavior or overconfidence in this area. Further towards the cluster boundary, the ambiguity in images seems to increase, as the CSF is able to detect the failures (light blue layer of dots). A layer of “false alarms” follows (brown colored dots), where decisions are correct, but confidence is still low.\\\\
        \textbf{SF-Visuals generates insights across tasks and distribution shifts.} \textbf{i.i.d. (no shift):} This analysis reveals how simple class clustering (no distribution shifts involved) can help to gain intuition on the most severe silent failures (examples selected as the two highest-confidence failures). On the lung nodule CT data (Figure~\ref{fig:bottom_up}a), we see how the classifier and CSF break down when a malignant sample (typically: small bright, round) exhibits characteristics typical to benign lesions (larger, less cohesive contour, darker) and vice versa. This pattern of contrary class characteristics is also observed on the dermoscopy dataset (\ref{fig:bottom_up}c). The failure example at the top is particularly severe, and localization in the scatter plot reveals a position deep inside the 'benign' cluster indicating either a severe sampling error in the dataset (e.g. underrepresented lesion sub-type) or simply a wrong label.
        \textbf{Corruption shift:} Figures~\ref{fig:bottom_up}b and \ref{fig:bottom_up}d show for the Lung Nodule CT data and the dermoscopy data, respectively, how corruptions can lead to silent failures in low-confident predictions. In both examples, the brightening of the image leads to a malignant lesion taking on benign characteristics (brighter and smoother skin on the dermoscopy data, decreased contrast between lesion and background on the Lung Nodule CT data).
        \textbf{Acquisition shift:} Additionally to the example in Figure~\ref{fig:overview}, Figure~\ref{fig:bottom_up}e shows how the proposed tool visualizes an acquisition shift on the chest X-ray data. While this reveals an increased blurriness in the target domain, it is difficult to derive further insights involving specific pathologies without a clinical expert. Figure~\ref{fig:bottom_up}h shows a classification failure from a target clinical center together with the model's confidence as measured by MSR and DG. While MSR assigns the prediction low confidence thereby catching the failure, DG assigns high confidence for the same model and prediction, causing a silent failure. This example shows how the tool allows the comparison of CSFs and can help to identify failure modes specific to each CSF.
        \textbf{Manifestation shift:} On the dermoscopy data (Figure~\ref{fig:bottom_up}g), we see how a manifestation shift can cause silent failures. The benign lesions in the target domain are similar to the malignant lesions in the source domain (rough skin, irregular shapes), and indeed the two failures in the target domain seem to fall into this trap. On the lung nodule CT data ((Figure~\ref{fig:bottom_up}f), we observe a  visual distinction between the spiculated target domain (spiked surface) and the non-spiculated source domain (smooth surface).

\section{Conclusion} 
    \label{sec:conclusion}
    We see two major opportunities for this work to make an impact on the community. 1) We hope the revealed shortcomings of current systems on biomedical tasks in combination with the deeper understanding of CSF behaviors granted by SF-Visuals will catalyze research towards a new generation of more reliable CSFs. 2) This study shows that in order to progress towards reliable ML systems, a deeper understanding of the data itself is required. SF-Visuals can help to bridge this gap and equip researchers with a better intuition of when and how to employ ML systems for a particular task.

\section*{Acknowledgements}
This work was funded by Helmholtz Imaging (HI), a platform of the Helmholtz Incubator on Information and Data Science.

\bibliographystyle{splncs04}
\bibliography{references.bib}

\begin{thebibliography}{10}
\providecommand{\url}[1]{\texttt{#1}}
\providecommand{\urlprefix}{URL }
\providecommand{\doi}[1]{https://doi.org/#1}

\bibitem{armatoLungImageDatabase2011}
Armato, S.G., McLennan, G., Bidaut, L., {McNitt-Gray}, M.F., Meyer, C.R.,
  et~al.: The {{Lung Image Database Consortium}} ({{LIDC}}) and {{Image
  Database Resource Initiative}} ({{IDRI}}): {{A Completed Reference Database}}
  of {{Lung Nodules}} on {{CT Scans}}: {{The LIDC}}/{{IDRI}} thoracic {{CT}}
  database of lung nodules. Medical Physics  \textbf{38}(2),  915--931 (Jan
  2011). \doi{10.1118/1.3528204}

\bibitem{bandBenchmarkingBayesianDeep2022}
Band, N., Rudner, T.G.J., Feng, Q., Filos, A., Nado, Z., et~al.: Benchmarking
  {{Bayesian Deep Learning}} on {{Diabetic Retinopathy Detection Tasks}}. In:
  Thirty-Fifth {{Conference}} on {{Neural Information Processing Systems
  Datasets}} and {{Benchmarks Track}} ({{Round}} 2) (Jan 2022)

\bibitem{bernhardtFailureDetectionMedical2022}
Bernhardt, M., Ribeiro, F.D.S., Glocker, B.: Failure {{Detection}} in {{Medical
  Image Classification}}: {{A Reality Check}} and {{Benchmarking Testbed}} (Oct
  2022). \doi{10.48550/arXiv.2205.14094}

\bibitem{castroCausalityMattersMedical2020a}
Castro, D.C., Walker, I., Glocker, B.: Causality matters in medical imaging.
  Nature Communications  \textbf{11}(1), ~3673 (Jul 2020).
  \doi{10.1038/s41467-020-17478-w}

\bibitem{corbiereAddressingFailurePrediction2019a}
Corbi{\`e}re, C., THOME, N., {Bar-Hen}, A., Cord, M., P{\'e}rez, P.: Addressing
  {{Failure Prediction}} by {{Learning Model Confidence}}. In: NeurIPS.
  vol.~32. {Curran Associates, Inc.} (2019)

\bibitem{devriesLearningConfidenceOutofDistribution2018a}
DeVries, T., Taylor, G.W.: Learning {{Confidence}} for {{Out-of-Distribution
  Detection}} in {{Neural Networks}} (Feb 2018).
  \doi{10.48550/arXiv.1802.04865}

\bibitem{fortExploringLimitsOutofDistribution2021}
Fort, S., Ren, J., Lakshminarayanan, B.: Exploring the {{Limits}} of
  {{Out-of-Distribution Detection}}. arXiv:2106.03004 [cs]  (Jul 2021)

\bibitem{galDropoutBayesianApproximation2016}
Gal, Y., Ghahramani, Z.: Dropout as a {{Bayesian Approximation}}:
  {{Representing Model Uncertainty}} in {{Deep Learning}}. In: ICML. pp.
  1050--1059. {PMLR} (Jun 2016)

\bibitem{geifmanSelectiveClassificationDeep2017}
Geifman, Y., {El-Yaniv}, R.: Selective {{Classification}} for {{Deep Neural
  Networks}}. arXiv:1705.08500 [cs]  (Jun 2017)

\bibitem{geifmanSelectiveNetDeepNeural2019}
Geifman, Y., {El-Yaniv}, R.: {{SelectiveNet}}: {{A Deep Neural Network}} with
  an {{Integrated Reject Option}}. arXiv:1901.09192 [cs, stat]  (Jun 2019)

\bibitem{hendrycksBaselineDetectingMisclassified2018}
Hendrycks, D., Gimpel, K.: A {{Baseline}} for {{Detecting Misclassified}} and
  {{Out-of-Distribution Examples}} in {{Neural Networks}}. arXiv:1610.02136
  [cs]  (Oct 2018)

\bibitem{huangDenselyConnectedConvolutional2017}
Huang, G., Liu, Z., {van der Maaten}, L., Weinberger, K.Q.: Densely {{Connected
  Convolutional Networks}}. In: CVPR. pp. 4700--4708 (2017)

\bibitem{idrissiImageNetXUnderstandingModel2022}
Idrissi, B.Y., Bouchacourt, D., Balestriero, R., Evtimov, I., Hazirbas, C.,
  et~al.: {{ImageNet-X}}: {{Understanding Model Mistakes}} with {{Factor}} of
  {{Variation Annotations}} (Nov 2022). \doi{10.48550/arXiv.2211.01866}

\bibitem{irvinCheXpertLargeChest2019}
Irvin, J., Rajpurkar, P., Ko, M., Yu, Y., {Ciurea-Ilcus}, S., et~al.:
  {{CheXpert}}: {{A Large Chest Radiograph Dataset}} with {{Uncertainty
  Labels}} and {{Expert Comparison}} (Jan 2019).
  \doi{10.48550/arXiv.1901.07031}

\bibitem{jaegerCallReflectEvaluation2022a}
Jaeger, P.F., L{\"u}th, C.T., Klein, L., Bungert, T.J.: A call to reflect on
  evaluation practices for failure detection in image classification. In:
  International Conference on Learning Representations (2023),
  \url{https://openreview.net/forum?id=YnkGMIh0gvX}

\bibitem{johnsonMIMICIIIFreelyAccessible2016}
Johnson, A.E.W., Pollard, T.J., Shen, L., Lehman, L.w.H., Feng, M., et~al.:
  {{MIMIC-III}}, a freely accessible critical care database. Scientific Data
  \textbf{3}(1),  160035 (May 2016). \doi{10.1038/sdata.2016.35}

\bibitem{kawaharaSevenPointChecklistSkin2019}
Kawahara, J., Daneshvar, S., Argenziano, G., Hamarneh, G.: Seven-{{Point
  Checklist}} and {{Skin Lesion Classification Using Multitask Multimodal
  Neural Nets}}. IEEE Journal of Biomedical and Health Informatics
  \textbf{23}(2),  538--546 (Mar 2019). \doi{10.1109/JBHI.2018.2824327}

\bibitem{kendallWhatUncertaintiesWe2017}
Kendall, A., Gal, Y.: What {{Uncertainties Do We Need}} in {{Bayesian Deep
  Learning}} for {{Computer Vision}}? arXiv:1703.04977 [cs]  (Mar 2017)

\bibitem{kohWILDSBenchmarkIntheWild2021}
Koh, P.W., Sagawa, S., Marklund, H., Xie, S.M., Zhang, M., et~al.: {{WILDS}}:
  {{A Benchmark}} of in-the-{{Wild Distribution Shifts}} (Jul 2021).
  \doi{10.48550/arXiv.2012.07421}

\bibitem{leeSimpleUnifiedFramework2018}
Lee, K., Lee, K., Lee, H., Shin, J.: A {{Simple Unified Framework}} for
  {{Detecting Out-of-Distribution Samples}} and {{Adversarial Attacks}}. In:
  NeurIPS. vol.~31. {Curran Associates, Inc.} (2018)

\bibitem{liangEnhancingReliabilityOutofdistribution2020}
Liang, S., Li, Y., Srikant, R.: Enhancing {{The Reliability}} of
  {{Out-of-distribution Image Detection}} in {{Neural Networks}}.
  arXiv:1706.02690 [cs, stat]  (Aug 2020)

\bibitem{liuDeepGamblersLearning2019}
Liu, Z., Wang, Z., Liang, P.P., Salakhutdinov, R.R., Morency, L.P., et~al.:
  Deep {{Gamblers}}: {{Learning}} to {{Abstain}} with {{Portfolio Theory}}. In:
  NeurIPS. vol.~32. {Curran Associates, Inc.} (2019)

\bibitem{malininPredictiveUncertaintyEstimation2018}
Malinin, A., Gales, M.: Predictive {{Uncertainty Estimation}} via {{Prior
  Networks}}. In: NeurIPS. vol.~31. {Curran Associates, Inc.} (2018)

\bibitem{mendoncaPH2DermoscopicImage2013}
Mendon{\c c}a, T., Ferreira, P.M., Marques, J.S., Marcal, A.R.S., Rozeira, J.:
  {{PH2}} - {{A}} dermoscopic image database for research and benchmarking. In:
  2013 35th {{Annual International Conference}} of the {{IEEE Engineering}} in
  {{Medicine}} and {{Biology Society}} ({{EMBC}}). pp. 5437--5440 (Jul 2013).
  \doi{10.1109/EMBC.2013.6610779}

\bibitem{ovadiaCanYouTrust2019}
Ovadia, Y., Fertig, E., Ren, J., Nado, Z., Sculley, D., et~al.: Can you trust
  your model' s uncertainty? {{Evaluating}} predictive uncertainty under
  dataset shift. In: NeurIPS. vol.~32. {Curran Associates, Inc.} (2019)

\bibitem{rotembergPatientcentricDatasetImages2021}
Rotemberg, V., Kurtansky, N., {Betz-Stablein}, B., Caffery, L., Chousakos, E.,
  et~al.: A patient-centric dataset of images and metadata for identifying
  melanomas using clinical context. Scientific Data  \textbf{8}(1), ~34 (Jan
  2021). \doi{10.1038/s41597-021-00815-z}

\bibitem{santurkarBREEDSBenchmarksSubpopulation2022}
Santurkar, S., Tsipras, D., Madry, A.: {{BREEDS}}: {{Benchmarks}} for
  {{Subpopulation Shift}}. In: International {{Conference}} on {{Learning
  Representations}} (Feb 2022)

\bibitem{sypetkowskiRxRx1DatasetEvaluating2023}
Sypetkowski, M., Rezanejad, M., Saberian, S., Kraus, O., Urbanik, J., et~al.:
  {{RxRx1}}: {{A Dataset}} for {{Evaluating Experimental Batch Correction
  Methods}} (Jan 2023). \doi{10.48550/arXiv.2301.05768}

\bibitem{tanEfficientNetRethinkingModel2019}
Tan, M., Le, Q.: {{EfficientNet}}: {{Rethinking Model Scaling}} for
  {{Convolutional Neural Networks}}. In: ICML. pp. 6105--6114. {PMLR} (May
  2019)

\bibitem{tschandlHAM10000DatasetLarge2018}
Tschandl, P., Rosendahl, C., Kittler, H.: The {{HAM10000}} dataset, a large
  collection of multi-source dermatoscopic images of common pigmented skin
  lesions. Scientific Data  \textbf{5}(1),  180161 (Aug 2018).
  \doi{10.1038/sdata.2018.161}

\bibitem{wangChestXRay8HospitalScaleChest2017}
Wang, X., Peng, Y., Lu, L., Lu, Z., Bagheri, M., et~al.: {{ChestX-Ray8}}:
  {{Hospital-Scale Chest X-Ray Database}} and {{Benchmarks}} on
  {{Weakly-Supervised Classification}} and {{Localization}} of {{Common Thorax
  Diseases}}. In: CVPR. pp. 3462--3471 (Jul 2017). \doi{10.1109/CVPR.2017.369}

\bibitem{winkensContrastiveTrainingImproved2020}
Winkens, J., Bunel, R., Roy, A.G., Stanforth, R., Natarajan, V., et~al.:
  Contrastive {{Training}} for {{Improved Out-of-Distribution Detection}}.
  arXiv:2007.05566 [cs, stat]  (Jul 2020)

\bibitem{zhangBenchmarkingRobustnessDeep2022}
Zhang, Y., Sun, Y., Li, H., Zheng, S., Zhu, C., et~al.: Benchmarking
  the~{{Robustness}} of~{{Deep Neural Networks}} to~{{Common Corruptions}}
  in~{{Digital Pathology}}. In: Wang, L., Dou, Q., Fletcher, P.T., Speidel, S.,
  Li, S. (eds.) MICCAI. pp. 242--252. Lecture {{Notes}} in {{Computer
  Science}}, {Springer Nature Switzerland}, {Cham} (2022).
  \doi{10.1007/978-3-031-16434-7\_24}

\end{thebibliography}

\end{document}